\documentclass[twocolumn,showpacs,preprintnumbers,amsmath,amssymb]{revtex4}

\usepackage{graphicx}
\usepackage{dcolumn}
\usepackage{bm}

\begin{document}

\preprint{APS/123-QED}

\title{Direct Experimental Evidence of Exciton-Phonon Bound States in Carbon Nanotubes}

\author{Fl\'{a}vio Plentz}
\email[]{plentz@fisica.ufmg.br}

\author{Henrique B. Ribeiro}
\author{Ado Jorio}
\author{Marcos A. Pimenta}

\affiliation{Departmento de F{\'i}sica, Universidade Federal de Minas Gerais, Belo Horizonte, MG, 30123-970 Brazil}

\author{Michael S. Strano}%
\affiliation{Department of Chemistry and Department of Chemical and Biomolecular Engineering,
University of Illinois at Urbana-Champaign, Urbana, Illinois 61801}

\begin{abstract}

We present direct experimental observation of exciton-phonon bound states in the photoluminescence excitation spectra of isolated single walled carbon nanotubes in aqueous suspension. The photoluminescence excitation spectra from several distinct single-walled carbon nanotubes show the presence of at least one sideband related to the tangential modes, lying {200 meV} above the main absorption/emission peak. Both the energy position and line shapes of the sidebands are in excellent agreement with recent calculations [PRL {\bf 94},027402 (2005)] that predict the existence of exciton-phonon bound states, a sizable spectral weight transfer to these exciton-phonon complexes and that the amount of this transfer depends on the specific nanotube structure and diameter. The observation of these novel exciton-phonon complexes is a strong indication that the optical properties of carbon nanotubes have an excitonic nature and also of the central role played by phonons in describing the excitation and recombination mechanisms in carbon nanotubes.

\end{abstract}

\maketitle

Single-walled carbon nanotubes (SWNT) exhibit optical properties that clearly show the one-dimensional character of their electronic structure \cite{jorio1}, allowing for the observation of Photoluminescence \cite{LefebvrePRB04}, Raman \cite{JorioPRL01} and Rayleigh scattering \cite{BrussScience04} at the single carbon nanotube level. Therefore, SWNTs are suitable for the investigation of fundamental physics phenomena in one-dimensional systems. Carbon nanotubes are also promising materials for applications in nanoelectronic and nanophotonic devices \cite{Avouris1}.

The nature of the optical transitions responsible for the optical properties in SWNTs has been the object of intense debate and, in the light of the possible application of SWNTs to nanophotonic devices, acquired a remarkable importance. The majority of authors have discussed their experimental observations in terms of band-band transitions involving free electron-hole pairs, while recent calculations show that the excitonic effect in carbon nanotubes is very strong, and that their photophysics are dictated by exciton states \cite{spataruPRL04}. From the experimental standpoint, a series of ``anomalous'' resonances in the photoluminescence excitation spectra (PLE) spectra of SWNT-DNA hybrids have been observed and discussed in terms of phonon mediated absorption and recombination channels \cite{grace1}, bringing into evidence the important role of electron-phonon interactions in the description of the SWNT optical properties. On Ref. \onlinecite{grace1} it was assumed that the optical transitions had an excitonic character, as suggested by recent theoretical calculations \cite{spataruPRL04, theoExciton2}, and that the observed resonances were due to several mechanisms involving the creation, annihilation and scattering of excitons and phonons. Similar observations were made on isolated SWNT and discussed within the framework of free electron-hole pair (band-band) transitions \cite{htoon1}. The discussion of most experimental observations using different models (exciton {\it vs.} band-band) has been successful due to the one-dimensional nature of the nanotubes -- leading to sharp one-dimensional van Hove singularities in the electronic density of states -- and to the close energy cancellation between many-body and excitonic corrections to the free electron-hole picture \cite{spataruPRL04}. Therefore, in order clarify the character of the optical transitions, the sole analysis of the optical transitions energies is not conclusive. Recently, two-photon absorption has been used to investigate the character of the optical transitions in semiconducting SWNTs \cite{twophoton1, twophoton2} showing clear evidence of their excitonic nature.

Within the exciton picture, Perebeinos {\it et al.} \cite{perebeinosPRL05} predicted the existence of novel bound exciton-phonon complexes that should manifest as absorption sidebands to which sizable spectral weight from the exciton peaks are transfered. The energy position, intensity and line shape of these sidebands were calculated. They found a robust side-band related to tangential vibrations, appearing {200\,meV} above the main absorption peak, that is absent without exciton binding.

In this letter we report on the observation of absorption sidebands in the photoluminescence excitation spectra (PLE), for many different HiPco SWNTs on aqueous suspension, that are in excellent agreement with the picture predicted by Perebeinos {\it et al.} \cite{perebeinosPRL05}, thus providing strong experimental evidence for bound exciton-phonon states in SWNTs and also for the excitonic nature of optical transitions in carbon nanotubes.

The sample studied in this work is an aqueous suspension of SWNTs grown by the HiPco process \cite{willis} dispersed by sodium dodecyl sulfate (SDS) as surfactant \cite{Connell}. The sample was placed in a quartz cuvette with a {1 cm} optical path and excited with a tunable Ti-Sapphire laser (Spectra Physics Model 3900s) pumped by an Argon ion Laser (Spectra Physics Model 2075). The laser was focused onto the sample using a 10$\times$ Mitutoyo objective with a working distance of 30.5mm and the excitation energy range was between {1.20 eV} and {1.75 eV}.  The photoluminescence emission was collected in a back scattering configuration and dispersed by a SPEX 750M spectrometer equipped with a North Coast Germanium detector. The PL spectra was measured for 51 different excitation energies. Care has been taken in order to correct the measured spectra by the system spectral response. All spectra have been taken using a {20 mW} laser power.

\begin{figure}[!]
\includegraphics[scale=0.45]{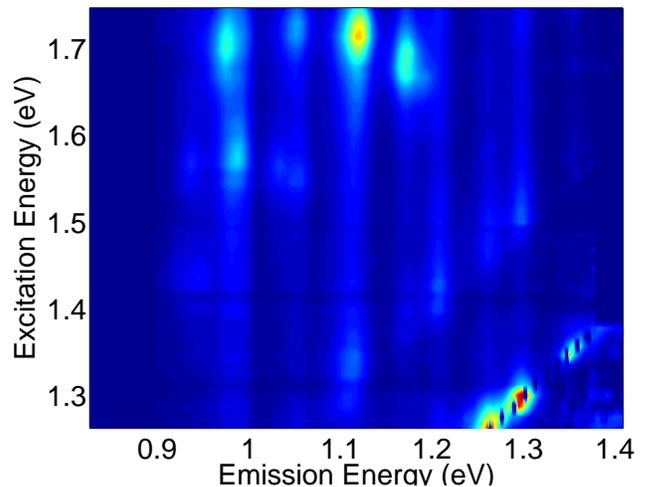}
\caption {(Color online) PLE map for SWNTs in SDS suspension measured with 51 laser lines between {1.20 eV} and {1.75 eV}. The stars indicate the position of some observed resonances.}
\label{f1}
\end{figure}

Figure \ref{f1} shows a 2D PLE plot obtained by measuring PLE spectra with 51 excitation energies, from {1.20 eV} through {1.75 eV}. In the PLE map the emission energies are on the horizontal axis and the excitation energies are on the vertical axis. Blue color represents low emission intensity while red indicates high intensity. Each vertical streak in the map represents an emission peak from a specific SWNT, and a high intensity spot in the map is associated to a resonance in the PLE spectra \cite{grace1, htoon1, Connell}. In this paper we label the radiative transitions as $E_{ii}^{j}$, where the subscript indexes {\it ii} ({\it i=1, 2, 3} etc) indicate the valence and conduction band band Van-Hove singularities (VHs) and the superscript {\it j} ({\it j=1s, 2s, 2p} etc) indicates the excitonic level. For instance, $E_{11}^{1s}$ and $E_{22}^{1s}$ correspond to the ground state of the exciton associated with the first and second VHs, respectively. For the excitation range used in this work we observe several resonances for which the excitation and recombination are both associated with the $E_{11}^{1s}$ transition. These resonances --- with the same excitation and recombination energy --- appear as a $45^{0}$ line at the lower right corner of the PLE map. Clear resonances, about {200 meV} above each $E_{11}^{1s}$--$E_{11}^{1s}$ resonance, can also be observed in the PLE map. 

\begin{figure}[!]
\includegraphics[scale=0.29,angle=-90]{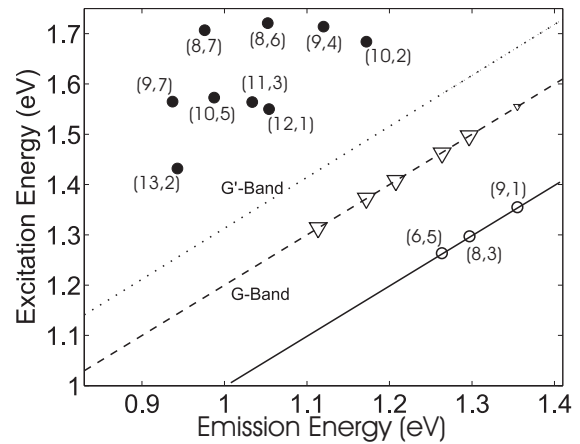}
\caption {(Color online) Main features observed in the PLE map. The black dots represent the usual resonances --- for which the excitation matches the $E_{22}^{1s}$ exciton --- and the open dots represent $E_{11}^{1s}$ resonances for which the excitation matches the $E_{11}^{1s}$ exciton. Each resonance is labeled --- by using the SWNT (n,m) indexes --- according to the SWNT responsible for the radiative transition. The open triagles represent the position of the resonaces associated with exciton-phonon bound states.}
\label{f2}
\end{figure}

In figure \ref{f2} we show schematically the main features observed in the PLE map. The black dots represent the usual resonances --- for which the excitation matches the $E_{22}^{1s}$ exciton --- and the open dots represent $E_{11}^{1s}$ resonances for which the excitation matches the $E_{11}^{1s}$ exciton. Each resonance is labeled --- by using the SWNT (n,m) indexes --- according to the SWNT responsible for the radiative transition. It is worth noticing that the assignments have been made based on a band-band model \cite{bachilo}. Since there is a near cancellation between many-body and excitonic corrections and the assignments made on Ref. \onlinecite{bachilo} also reflect the geometry of the graphene lattice, the results can also be used in our case. The open triangles are associated enhancements in the PL intensity that cannot be related to the usual $E_{ii}^{j}$ resonances. A solid line connecting the open dots and a dashed line connecting the open triangles have been drawn as a guide to the eyes. The line connecting the open triangles is about 200 meV above the line associated with the $E_{11}^{1s}$ resonances, an energy difference that corresponds to the Raman G band \cite{ReviewMillie}. These features, that show up as side-bands in the PLE spectra, can not be understood in terms of a model that considers only excitonic or band-band transitions. We also draw a dotted line, about 330 meV above the open dots, to stress weaker enhancements in the PL intensity that can be perceived in the PLE map. This energy difference corresponds to G' band in the Raman spectra of SWNTs. Since the observed sidebands are located above the $E_{11}^{1s}$ resonances by an energy that corresponds to the SWNT Raman G band, it is reasonable to assume that this absorption process involves the participation the vibrational modes of the SWNTs.

\begin{figure}[!]
\includegraphics[scale=.38, angle=0]{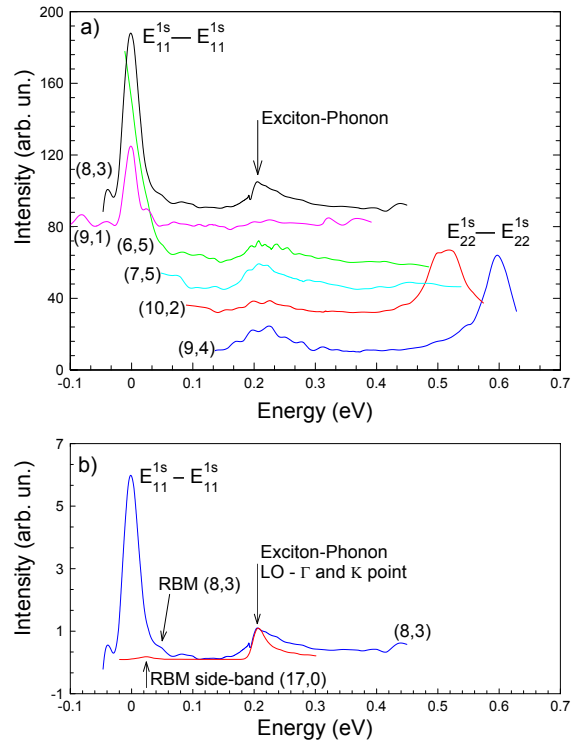}
\caption {(Color online)Figure 3a shows the PLE profiles for several SWNTs exhibiting a sideband 200 meV above the $E_{11}^{1s}$ resonance. The origin of this sideband is the resonance absorption of light when the photon energy matches energy of the exciton-phonon bound state. Fig 3b shows the PLE profile for the (8,3) SWNT together with the calculated spectra for the (17,0) SWNT taken from Ref. \onlinecite{perebeinosPRL05}}
\label{f3}
\end{figure}

In the present work, we measured and carry out a line shape analysis of the PLE profile. The absorption profile is obtained by performing a vertical cut in the PLE map at energies that correspond to the emission of a given nanotube. For some SWNTs -- (8,3) and (9,1) -- both the $E_{11}^{1s}$ resonance and phonon related sidebands are present. In figure \ref{f3}a we show the plot of such profiles for SWNTs for which the range of the excitation energies allow for the observation of the sidebands. For each tube, the zero energy is set at the energy of the $E_{11}^{1s}$ exciton. It can be clearly seen that all tubes show a sideband, with similar line shape, about {200 meV} above the fundamental optical gap. Since this energy difference corresponds to the SWNT Raman G band phonon, there is the possibility that this band could be related to a resonance Raman process where the scattered photon would be in resonance with the energy of the $E_{11}^{1s}$ exciton since, in this case, the incoming photon has an energy $\hbar\omega_{in}=\hbar\omega_{out}+\hbar\omega_{Gphonon}$ \cite{grace1, htoon1}. This possibility is ruled out due to the line shapes of the PLE profile and the emission spectra. A Raman process would be represented in the emission spectra --- obtained by taking a horizontal cut in the PLE map --- by a much narrower, Lorentzian line shape and its resonance profile would not have a broad line shape enlogated towards the high energy side. We attribute the observed side-bands to a resonance whose origin is the absorption of light to a bound exciton-phonon state as proposed by Perebeinos $\it{et al.}$ \cite{perebeinosPRL05}.

In the exciton-phonon complex, the zero momentum exciton is mixed with excitons of finite momentum $\roarrow q$ and phonons with momentum $-\roarrow q$ through electron-phonon interaction, conserving momentum and leading to a bound state that can be populated by absorbing a photon with suitable energy. The main contribution for the exciton-phonon bound state comes from optical LO phonon branch at $\bf K$ and $\bf \Gamma$ points of the graphene Brillouin zone giving rise to an absorption band about {200 meV} above the $E_{11}^{1s}$ exciton absorption. A significant fraction of the spectral weight is transferred from the exciton peak to the exciton-phonon complex. In figure \ref{f3}b we show the PLE profile for the (8,3) SWNT and the calculated spectra for the (17,0) that has been taken from Ref. \onlinecite{perebeinosPRL05}. The intensity of the calculated exciton-phonon sideband has been renormalized to match the intensity of the (8,3) SWNT sideband. The agreement is remarkable. Since the energy position of this sideband depends mostly on the phonon energy it is expected a small diference between the (8,3) and (17,0) SWNTs since this particular sideband is associated with tangential vibrational modes (G band). There is also a small contribution from the radial breathing mode giving rise to an absorption band shifted towards the high energy side of the $E_{11}^{1s}$ peak by an energy corresponding aproximately to the RBM mode. The PLE profiles of the (6,5), (8,3) and (9,1) tubes show a shoulder in the high energy side of the $E_{11}^{1s}$ resonance about {40 meV} from the main peak. This energy corresponds roughly to the RMB mode energy but the signal to noise ratio in the data does not allow for a positive assignment. In figure \ref{f3}b the RBM sideband in the calculated spectra is downshifted with respect to the shoulder in the high energy side (8,3) SWNT $E_{11}^{1s}$ -- $E_{11}^{1s}$ resonance by aproximately {20 meV}, the expected energy shift between the RBM modes of the (17,0) and (8,3) SWNTs.

\begin{figure}[!]
\includegraphics[scale=.29,angle=0]{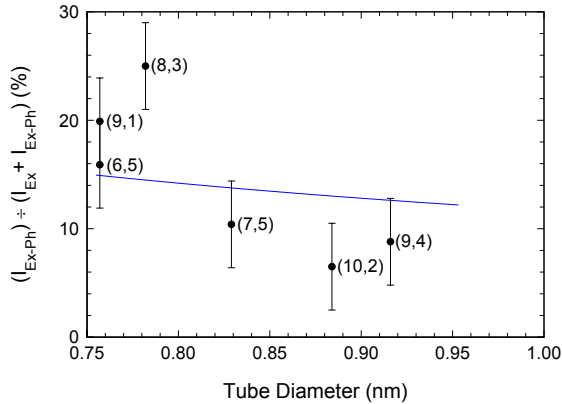}
\caption {(Color online) Diameter dependence of the spectral weight transfer from the exciton to the exciton-phonon sideband. The solid line is the theoretical curve for zig-zag SWNTs (Ref. \onlinecite{perebeinosPRL05}). There is a clear trend showing that larger diameter --- (7,5), (10,2) and (9,4) --- tubes have a lower spectral weight transfer than larger diameter tubes -- (6,5), (8,3) and (9,1), as predicted in the theory.} 
\label{f4}
\end{figure}

A diameter dependence of the spectral weight transfer from the exciton peak to the exciton-phonon is also predicted on Ref. \onlinecite{perebeinosPRL05}. The spectral weight transfer is shown to be larger for smaller diameter tubes with a $1/d_{t}$ dependence, where $d_{t}$ is the tube diameter. Figure \ref{f4} shows the measured spectral weight transfer together with the theoretical curve for zig-zag SWNTs taken from Ref. \onlinecite{perebeinosPRL05}. Since we measured the $E_{11}^{1s}$--$E_{11}^{1s}$ resonance intensity only for the (8,3) and (9,1) SWNTs, in order to calculate the spectral weight transfer ratios we used the relative intensity data for the $E_{11}^{1s}$--$E_{11}^{1s}$ resonances measured in a similar sample \cite{strano1} and renormalized by the integrated intensity of the $E_{11}^{1s}$ -- $E_{11}^{1s}$ for the (8,3) SWNT measured in our experiment. The qualitative agreement between the predicted and measured dependence is noticeable. There is a clear trend showing that larger diameter --- (7,5), (10,2) and (9,4) --- tubes have lower spectral weight transfer than larger diameter tubes -- (6,5), (8,3) and (9,1), and the absolute ($\agt 15\%$) is also consistent with the theory.

The experimental observation of these bound exciton-phonon states are a strong experimental evidence that the photophysics of SWNTs is dictated by excitons and that the strong interaction of the electronic and vibrational states are crutial in the description of the SWNTs optical properties and, as a consequence, in the operation and modeling of nanophotonic devices based on SWNTs. 

In carbon nanotubes the exciton charge density forms an electric dipole aligned along the tube axis and this fact promotes a strong coupling between the exciton and tangential phonons, giving rise to a bound state that can be viewed as a novel quasi-particle consisting of an exciton dressed by phonons. This effect should be general caracteristic of 1D systems with stable exciton states and sizeable exciton-phonon coupling. In fact, such an effect has also been shown to take place on polymers, in particular on PPV where the optical properties have been shown to be strongly influenced by excition-phonon coupling \cite{caldas1}. 

In summary, using PL and PLE spectroscopy we observe several $E_{11}^{1s}$-$E_{22}^{1s}$ and $E_{11}^{1s}$-$E_{11}^{1s}$ transitions and -- at energies between these two process -- resonances that cannot be undertood within the framework of band-band or purely excitonic transitions are also observed. By the line shape and the diameter dependence of the spectral weight transferred for the excitonic sideband, we were able to prove that an exciton-phonon coupling, leading to a bound-exciton state previously proposed by Perebeinos $\it et.al.$ \cite{perebeinosPRL05} are responsible for such strong absorption side-bands.  This is a clear experimental evidence that optical transitions in carbon nanotubes have an excitonic character and of the central role played by phonons in the description of the SWNTs optical properties.

\begin{acknowledgements}

We would like to thank R. Saito and for A. S. Ito for valuable
discussions. The Brazilian authors acknowledge financial support
from the Instituto de Nanosci\^encias - CNPq, CAPES and FAPEMIG.

\end{acknowledgements}

\bibliography{plpaper1}

\begin{thebibliography}{18}
\expandafter\ifx\csname natexlab\endcsname\relax\def\natexlab#1{#1}\fi
\expandafter\ifx\csname bibnamefont\endcsname\relax
  \def\bibnamefont#1{#1}\fi
\expandafter\ifx\csname bibfnamefont\endcsname\relax
  \def\bibfnamefont#1{#1}\fi
\expandafter\ifx\csname citenamefont\endcsname\relax
  \def\citenamefont#1{#1}\fi
\expandafter\ifx\csname url\endcsname\relax
  \def\url#1{\texttt{#1}}\fi
\expandafter\ifx\csname urlprefix\endcsname\relax\def\urlprefix{URL }\fi
\providecommand{\bibinfo}[2]{#2}
\providecommand{\eprint}[2][]{\url{#2}}

\bibitem[{\citenamefont{Jorio et~al.}(2004)\citenamefont{Jorio, Saito, Hertel,
  Weisman, Dresselhaus, and Dresselhaus}}]{jorio1}
\bibinfo{author}{\bibfnamefont{A.}~\bibnamefont{Jorio}},
  \bibinfo{author}{\bibfnamefont{R.}~\bibnamefont{Saito}},
  \bibinfo{author}{\bibfnamefont{T.}~\bibnamefont{Hertel}},
  \bibinfo{author}{\bibfnamefont{R.~B.} \bibnamefont{Weisman}},
  \bibinfo{author}{\bibfnamefont{G.}~\bibnamefont{Dresselhaus}},
  \bibnamefont{and} \bibinfo{author}{\bibfnamefont{M.~S.}
  \bibnamefont{Dresselhaus}}, \bibinfo{journal}{MRS Bulletin}
  \textbf{\bibinfo{volume}{29}}, \bibinfo{pages}{276} (\bibinfo{year}{2004}).

\bibitem[{\citenamefont{Lefebvre et~al.}(2004)\citenamefont{Lefebvre, Fraser,
  Finnie, , and Homma}}]{LefebvrePRB04}
\bibinfo{author}{\bibfnamefont{J.}~\bibnamefont{Lefebvre}},
  \bibinfo{author}{\bibfnamefont{J.~M.} \bibnamefont{Fraser}},
  \bibinfo{author}{\bibfnamefont{P.}~\bibnamefont{Finnie}}, , \bibnamefont{and}
  \bibinfo{author}{\bibfnamefont{Y.}~\bibnamefont{Homma}},
  \bibinfo{journal}{Phys Rev B} \textbf{\bibinfo{volume}{69}},
  \bibinfo{pages}{075403} (\bibinfo{year}{2004}).

\bibitem[{\citenamefont{Jorio et~al.}(2001)\citenamefont{Jorio, Saito, Hfner,
  Lieber, Hunter, McClure, Dresselhaus, and Dresselhaus}}]{JorioPRL01}
\bibinfo{author}{\bibfnamefont{A.}~\bibnamefont{Jorio}},
  \bibinfo{author}{\bibfnamefont{R.}~\bibnamefont{Saito}},
  \bibinfo{author}{\bibfnamefont{J.~H.} \bibnamefont{Hfner}},
  \bibinfo{author}{\bibfnamefont{C.~M.} \bibnamefont{Lieber}},
  \bibinfo{author}{\bibfnamefont{M.}~\bibnamefont{Hunter}},
  \bibinfo{author}{\bibfnamefont{T.}~\bibnamefont{McClure}},
  \bibinfo{author}{\bibfnamefont{G.}~\bibnamefont{Dresselhaus}},
  \bibnamefont{and} \bibinfo{author}{\bibfnamefont{M.~S.}
  \bibnamefont{Dresselhaus}}, \bibinfo{journal}{Phys. Rev. Lett.}
  \textbf{\bibinfo{volume}{86}}, \bibinfo{pages}{1118} (\bibinfo{year}{2001}).

\bibitem[{\citenamefont{Sfeir et~al.}(2004)\citenamefont{Sfeir, Wang, Huang,
  Chuang, Hone, O'Brien, Heinz, and Brus}}]{BrussScience04}
\bibinfo{author}{\bibfnamefont{M.~Y.} \bibnamefont{Sfeir}},
  \bibinfo{author}{\bibfnamefont{F.}~\bibnamefont{Wang}},
  \bibinfo{author}{\bibfnamefont{L.}~\bibnamefont{Huang}},
  \bibinfo{author}{\bibfnamefont{C.-C.} \bibnamefont{Chuang}},
  \bibinfo{author}{\bibfnamefont{J.}~\bibnamefont{Hone}},
  \bibinfo{author}{\bibfnamefont{S.~P.} \bibnamefont{O'Brien}},
  \bibinfo{author}{\bibfnamefont{T.~F.} \bibnamefont{Heinz}}, \bibnamefont{and}
  \bibinfo{author}{\bibfnamefont{L.~E.} \bibnamefont{Brus}},
  \bibinfo{journal}{Science} \textbf{\bibinfo{volume}{306}},
  \bibinfo{pages}{1540} (\bibinfo{year}{2004}).

\bibitem[{\citenamefont{Avouris}(2004)}]{Avouris1}
\bibinfo{author}{\bibfnamefont{P.}~\bibnamefont{Avouris}},
  \bibinfo{journal}{MRS Bulletin} \textbf{\bibinfo{volume}{29}},
  \bibinfo{pages}{403} (\bibinfo{year}{2004}).

\bibitem[{\citenamefont{Spataru et~al.}(2004)\citenamefont{Spataru,
  Ismail-Beigiand, Benedict, and Louie}}]{spataruPRL04}
\bibinfo{author}{\bibfnamefont{C.~D.} \bibnamefont{Spataru}},
  \bibinfo{author}{\bibfnamefont{S.}~\bibnamefont{Ismail-Beigiand}},
  \bibinfo{author}{\bibfnamefont{L.~X.} \bibnamefont{Benedict}},
  \bibnamefont{and} \bibinfo{author}{\bibfnamefont{S.~G.} \bibnamefont{Louie}},
  \bibinfo{journal}{Phys. Rev. Lett.} \textbf{\bibinfo{volume}{92}},
  \bibinfo{pages}{077402} (\bibinfo{year}{2004}).

\bibitem[{\citenamefont{Chou et~al.}(2005)\citenamefont{Chou, Plentz, Jiang,
  Saito, Nezich, Ribeiro, Jorio, Pimenta, Samsonidze, Santos et~al.}}]{grace1}
\bibinfo{author}{\bibfnamefont{S.~G.} \bibnamefont{Chou}},
  \bibinfo{author}{\bibfnamefont{F.}~\bibnamefont{Plentz}},
  \bibinfo{author}{\bibfnamefont{J.}~\bibnamefont{Jiang}},
  \bibinfo{author}{\bibfnamefont{R.}~\bibnamefont{Saito}},
  \bibinfo{author}{\bibfnamefont{D.}~\bibnamefont{Nezich}},
  \bibinfo{author}{\bibfnamefont{H.~B.} \bibnamefont{Ribeiro}},
  \bibinfo{author}{\bibfnamefont{A.}~\bibnamefont{Jorio}},
  \bibinfo{author}{\bibfnamefont{M.~A.} \bibnamefont{Pimenta}},
  \bibinfo{author}{\bibfnamefont{G.~G.} \bibnamefont{Samsonidze}},
  \bibinfo{author}{\bibfnamefont{A.~P.} \bibnamefont{Santos}},
  \bibnamefont{et~al.}, \bibinfo{journal}{Phys. Rev. Lett.}
  \textbf{\bibinfo{volume}{94}}, \bibinfo{pages}{127402}
  (\bibinfo{year}{2005}).

\bibitem[{\citenamefont{Perebeinos et~al.}(2004)\citenamefont{Perebeinos,
  Tersoff, and Avouris}}]{theoExciton2}
\bibinfo{author}{\bibfnamefont{V.}~\bibnamefont{Perebeinos}},
  \bibinfo{author}{\bibfnamefont{J.}~\bibnamefont{Tersoff}}, \bibnamefont{and}
  \bibinfo{author}{\bibfnamefont{P.}~\bibnamefont{Avouris}},
  \bibinfo{journal}{Phys. Rev. Lett.} \textbf{\bibinfo{volume}{92}},
  \bibinfo{pages}{257402} (\bibinfo{year}{2004}).

\bibitem[{\citenamefont{Htoon et~al.}(2005)\citenamefont{Htoon, O'Connell,
  Doorn, and Klimov}}]{htoon1}
\bibinfo{author}{\bibfnamefont{H.}~\bibnamefont{Htoon}},
  \bibinfo{author}{\bibfnamefont{M.~J.} \bibnamefont{O'Connell}},
  \bibinfo{author}{\bibfnamefont{S.~K.} \bibnamefont{Doorn}}, \bibnamefont{and}
  \bibinfo{author}{\bibfnamefont{V.~I.} \bibnamefont{Klimov}},
  \bibinfo{journal}{Phys. Rev. Lett.} \textbf{\bibinfo{volume}{94}},
  \bibinfo{pages}{127403} (\bibinfo{year}{2005}).

\bibitem[{\citenamefont{Wang et~al.}(2005)\citenamefont{Wang, Dukovic, Brus,
  and Heinz}}]{twophoton1}
\bibinfo{author}{\bibfnamefont{F.}~\bibnamefont{Wang}},
  \bibinfo{author}{\bibfnamefont{G.}~\bibnamefont{Dukovic}},
  \bibinfo{author}{\bibfnamefont{L.~E.} \bibnamefont{Brus}}, \bibnamefont{and}
  \bibinfo{author}{\bibfnamefont{T.~F.} \bibnamefont{Heinz}},
  \bibinfo{journal}{Science} \textbf{\bibinfo{volume}{308}},
  \bibinfo{pages}{838} (\bibinfo{year}{2005}).

\bibitem[{\citenamefont{Maultzsch et~al.}(2005)\citenamefont{Maultzsch,
  Pomraenke, Reich, Chang, Prezzi, Ruini, Molinari, Strano, Thomsen, and
  Lienau}}]{twophoton2}
\bibinfo{author}{\bibfnamefont{J.}~\bibnamefont{Maultzsch}},
  \bibinfo{author}{\bibfnamefont{R.}~\bibnamefont{Pomraenke}},
  \bibinfo{author}{\bibfnamefont{S.}~\bibnamefont{Reich}},
  \bibinfo{author}{\bibfnamefont{E.}~\bibnamefont{Chang}},
  \bibinfo{author}{\bibfnamefont{D.}~\bibnamefont{Prezzi}},
  \bibinfo{author}{\bibfnamefont{A.}~\bibnamefont{Ruini}},
  \bibinfo{author}{\bibfnamefont{E.}~\bibnamefont{Molinari}},
  \bibinfo{author}{\bibfnamefont{M.~S.} \bibnamefont{Strano}},
  \bibinfo{author}{\bibfnamefont{C.}~\bibnamefont{Thomsen}}, \bibnamefont{and}
  \bibinfo{author}{\bibfnamefont{C.}~\bibnamefont{Lienau}},
  \bibinfo{journal}{cond-mat/0505150}  (\bibinfo{year}{2005}).

\bibitem[{\citenamefont{Perebeinos et~al.}(2005)\citenamefont{Perebeinos,
  Tersoff, and Avouris}}]{perebeinosPRL05}
\bibinfo{author}{\bibfnamefont{V.}~\bibnamefont{Perebeinos}},
  \bibinfo{author}{\bibfnamefont{J.}~\bibnamefont{Tersoff}}, \bibnamefont{and}
  \bibinfo{author}{\bibfnamefont{P.}~\bibnamefont{Avouris}},
  \bibinfo{journal}{Phys. Rev. Lett.} \textbf{\bibinfo{volume}{94}},
  \bibinfo{pages}{027402} (\bibinfo{year}{2005}).

\bibitem[{\citenamefont{Bronikowski et~al.}(2001)\citenamefont{Bronikowski,
  Willis, Colbert, Smith, and Smalley}}]{willis}
\bibinfo{author}{\bibfnamefont{M.~J.} \bibnamefont{Bronikowski}},
  \bibinfo{author}{\bibfnamefont{P.~A.} \bibnamefont{Willis}},
  \bibinfo{author}{\bibfnamefont{D.~T.} \bibnamefont{Colbert}},
  \bibinfo{author}{\bibfnamefont{K.~A.} \bibnamefont{Smith}}, \bibnamefont{and}
  \bibinfo{author}{\bibfnamefont{R.~E.} \bibnamefont{Smalley}},
  \bibinfo{journal}{J. Vacuum Sci. Technol. A} \textbf{\bibinfo{volume}{19}},
  \bibinfo{pages}{1800} (\bibinfo{year}{2001}).

\bibitem[{\citenamefont{O'Connell et~al.}(2002)\citenamefont{O'Connell,
  Bachilo, Huffman, Moore, Strano, Haroz, Rialon, Boul, Noon, Kittrell
  et~al.}}]{Connell}
\bibinfo{author}{\bibfnamefont{M.~J.} \bibnamefont{O'Connell}},
  \bibinfo{author}{\bibfnamefont{S.~M.} \bibnamefont{Bachilo}},
  \bibinfo{author}{\bibfnamefont{C.~B.} \bibnamefont{Huffman}},
  \bibinfo{author}{\bibfnamefont{V.~C.} \bibnamefont{Moore}},
  \bibinfo{author}{\bibfnamefont{M.~S.} \bibnamefont{Strano}},
  \bibinfo{author}{\bibfnamefont{E.~H.} \bibnamefont{Haroz}},
  \bibinfo{author}{\bibfnamefont{K.~L.} \bibnamefont{Rialon}},
  \bibinfo{author}{\bibfnamefont{P.~J.} \bibnamefont{Boul}},
  \bibinfo{author}{\bibfnamefont{W.~H.} \bibnamefont{Noon}},
  \bibinfo{author}{\bibfnamefont{C.}~\bibnamefont{Kittrell}},
  \bibnamefont{et~al.}, \bibinfo{journal}{Science}
  \textbf{\bibinfo{volume}{297}}, \bibinfo{pages}{593} (\bibinfo{year}{2002}).

\bibitem[{\citenamefont{Bachilo et~al.}(2002)\citenamefont{Bachilo, Strano,
  Kittrell, Hauge, Smalley, and Weisman}}]{bachilo}
\bibinfo{author}{\bibfnamefont{S.~M.} \bibnamefont{Bachilo}},
  \bibinfo{author}{\bibfnamefont{M.~S.} \bibnamefont{Strano}},
  \bibinfo{author}{\bibfnamefont{C.}~\bibnamefont{Kittrell}},
  \bibinfo{author}{\bibfnamefont{R.~H.} \bibnamefont{Hauge}},
  \bibinfo{author}{\bibfnamefont{R.~E.} \bibnamefont{Smalley}},
  \bibnamefont{and} \bibinfo{author}{\bibfnamefont{R.~B.}
  \bibnamefont{Weisman}}, \bibinfo{journal}{Science}
  \textbf{\bibinfo{volume}{298}}, \bibinfo{pages}{2361} (\bibinfo{year}{2002}).

\bibitem[{\citenamefont{Dresselhaus et~al.}(2005)\citenamefont{Dresselhaus,
  Dresselhaus, Saito, and Jorio}}]{ReviewMillie}
\bibinfo{author}{\bibfnamefont{M.~S.} \bibnamefont{Dresselhaus}},
  \bibinfo{author}{\bibfnamefont{G.}~\bibnamefont{Dresselhaus}},
  \bibinfo{author}{\bibfnamefont{R.}~\bibnamefont{Saito}}, \bibnamefont{and}
  \bibinfo{author}{\bibfnamefont{A.}~\bibnamefont{Jorio}},
  \bibinfo{journal}{Physics Reports} \textbf{\bibinfo{volume}{409}},
  \bibinfo{pages}{47} (\bibinfo{year}{2005}).

\bibitem[{\citenamefont{Strano}()}]{strano1}
\bibinfo{author}{\bibfnamefont{M.~S.} \bibnamefont{Strano}},
  \eprint{unpublished}.

\bibitem[{\citenamefont{Ruini et~al.}(2002)\citenamefont{Ruini, Caldas, Bussi,
  and Molinari}}]{caldas1}
\bibinfo{author}{\bibfnamefont{A.}~\bibnamefont{Ruini}},
  \bibinfo{author}{\bibfnamefont{M.~J.} \bibnamefont{Caldas}},
  \bibinfo{author}{\bibfnamefont{G.}~\bibnamefont{Bussi}}, \bibnamefont{and}
  \bibinfo{author}{\bibfnamefont{E.}~\bibnamefont{Molinari}},
  \bibinfo{journal}{Phys. Rev. Lett.} \textbf{\bibinfo{volume}{88}},
  \bibinfo{pages}{206403} (\bibinfo{year}{2002}).

\end{thebibliography}

\end{document}